\documentclass{svproc}
\usepackage[
backend=biber,
style=alphabetic,
sorting=ynt
]{biblatex}

\usepackage{url}
\usepackage{listings}
\usepackage{graphicx}
\usepackage{amsmath}

\begin{document}

\mainmatter              
\title{Implementation of a framework for deploying AI inference engines in FPGAs}
\titlerunning{SLAC SNL Framework}  
%
\author{Ryan Herbst 
\and Ryan Coffee
\and Nathan Fronk 
\and Kukhee Kim
\and Kuktae Kim 
\and Larry Ruckman 
\and J.J. Russell}
\authorrunning{Ryan Herbst et al.} 
%
\tocauthor{Ryan Herbst, Ryan Coffee, Nathan Fronk, Kukhee Kim, Kuktae Kim, Larry Ruckman, J.J. Russell}
\institute{SLAC National Accelerator Laboratory, Menlo Park, CA 95024, USA,\\
\email{rherbst@slac.stanford.edu}}

\maketitle              

\begin{abstract}
The LCLS2 Free Electron Laser (FEL) will generate x-ray pulses to beamline experiments at up to 1Mhz. These experimentals will require new ultra-high rate (UHR) detectors that can operate at rates above 100 kHz and generate data throughputs upwards of 1 TB/s, a data velocity which requires prohibitively large investments in storage infrastructure. Machine Learning has demonstrated the potential to digest large datasets to extract relevant insights, however current implementations show latencies that are too high for real-time data reduction objectives. SLAC has endeavored on the creation of a software framework which translates MLs structures for deployment on Field Programmable Gate Arrays (FPGAs) deployed at the Edge of the data chain, close to the instrumentation. This framework leverages Xilinx’s HLS framework presenting an API modeled after the open source Keras interface to the TensorFlow library. This SLAC Neural Network Library (SNL) framework is designed with a streaming data approach, optimizing the data flow between layers, while minimizing the buffer data buffering requirements. The goal is to ensure the highest possible framerate while keeping the maximum latency constrained to the needs of the experiment. Our framework is designed to ensure the RTL implementation of the network layers supporting full re-deployment of weights and biases without requiring re-synthesis after training. The ability to reduce the precision of the implemented networks through quantization is necessary to optimize the use of both DSP and memory resources in the FPGA. We currently have a preliminary version of the toolset and are experimenting with both general purpose example networks and networks being designed for specific LCLS2 experiments. 
\end{abstract}

\section{Introduction}

New detectors for science and other applications have exponentially increased their pixel count and their frame rate, resulting in ever larger data rates. Some of these ultra-high rate (UHR) detectors can operate at rates above 100 kHz and generate data throughputs upwards of 1 TB/s, a data velocity which requires prohibitively large investments in storage infrastructure. \cite{osti_1630267} Machine Learning has demonstrated the potential to digest large datasets to extract relevant insights, however current implementations show latencies that are too high for real-time data reduction objectives. We intend to use machine learning inference models entirely deployed on a network of interconnected FPGAs allowing data to be pipelined for high throughput with ultra-low latency.

Edge Computing systems will receive the raw detector output and will preprocess, veto and classify the frame before sending the compressed information downstream for further analysis and/or storage. The re-programmability of FPGAs makes it possible to have a custom ML inference model for each detector and experiment. To facilitate the development and deployment of models for diverse experiments, we have created a framework (SNL) which will translate ML structures in FPGA code and deploy it to the FPGA network. 

\subsection{The SNL Framework}

The primary goal of SNL is to produce a high-performance, low latency FPGA implementation of an AI inference engine that can accommodate reasonably sized networks and be robust enough to adapt to changes when deployed in a real-time environment. A secondary goal is to make this as easy to use as possible without sacrificing those primary goals.

C++ templates used within the Xilinx Vitis HLS development environment and modeled after the Python Keras layer procedures were chosen as the implementation method to address the performance goals and the ease of use. Dynamic loading of weights and biases was chosen to achieve robustness by avoiding re-synthesizing the network when a new set of weights and biases were needed.  Once verified, deployment of a new set of weights and biases is procedurally the same as any other restart of the system.

The next two sections attempt to justify or at least explain the pluses and minuses of these decisions and how they help achieve the primary goal.

\subsection{Why C++ Templates?}

FPGAs work best the more that is statically known at compile/build time affording the compiler the best opportunity to optimize resources and latency/thru-put. This matches well with the target SNL application, \emph{e.g.} the topology of the inference engine is fixed, with known data sizes and loop iteration counts. Using C++ templates provides a mechanism to define this topology and, together with Xilinx HLS's palette of pragmas, allows the effective mapping of software concepts onto the FPGA resources.

The C++ templates are modeled as closely as possible in their form and function with the Python Keras layer methods. Given that a FPGA has a very different computational model from a CPU or GPU, there are necessarily differences.  The design goal was not to eliminate or hide these differences, but to limit their number to what was necessary to achieve the primary goals.  One of the important differences is the interface between layers is a streaming, not a memory interface. See section \ref{stream} 

This approach can be contrasted with what could be called a \emph{code that writes code} approach.  The following is not meant to be promoting one over the other. As with many things, one approach's strengths are the other's weaknesses. Users should pick the approach best suited to their problem and skill set.

In general, the \emph{code that writes code} approach is more turnkey and easier for users with minimal C++, Xilinx HLS and FPGA experience. The downside is that it is fairly rigid in its implementation and when things go wrong, even if it the user's mistake, it is often hard to track down the origin of the mistake. Even with good tools, the made-up names and layout of the generated code can become confusing and intractable. 

On the flip side, while care has been taken to make SNL as easy to use as possible, it does demand more expertise on the user's part. This is a deliberate design decision.  It is believed that the target application, AI at the edge, will be matched by users who have commensurate expertise in these areas. The hope is that by being just standard C++ code augmented with HLS pragmas, this gives the user greater control over the code and will allow greater performance, flexibility and the ability to track down errors when they inevitably occur. This flexibility is particularly useful when dealing with larger networks, for example allowing the user to trade performance with the finite FPGA resources.  

\subsection{Why Dynamic Loading Of Weights and Biases?}

There are two tactics one can take with the weights and biases that are calculated from the machine learning training. 
   \begin{itemize}
      \item{Build them into the code at synthesis time}
      \item{Load them at runtime}
   \end{itemize}
   
 Building the weights and biases into the code during the synthesis allows the compiler the very real and tangible opportunity to better optimize the code.  For example, weights that have little impact on the results may be pruned. It is noted that loading these at run-time eliminates this possible optimization and is at odds with the stated premium SNL places on performance. However, as in most engineering endeavors, there are trade-offs. There are two downsides to building the weights and biases into the FPGA image. Both involve operational time penalties
    \begin{itemize}
       \item{The time to re-synthesize the FPGA image}
       \item{The small, but not negligible, chance that the re-synthesis will fail}
    \end{itemize}
    
An assumption is that SNL's use will be in the high stakes real-time environment of running a facility or experimental data taking, where downtime is to be minimized. Presumably a new set of weights and biases is being deployed because changing conditions demand it, \emph{i.e} a set of new weights and biases must be deployed.

The first issue is just the reality that the time to re-synthesize networks for an FPGA can run into the multiple hours.  This time is somewhat predictable and generally accepted as just the cost of using FPGAs. Said another way, it can be properly factored into the operational and scheduling,

The second entertains the possibility of the re-synthesis failing. An example of such a failure is if the previous set of weights was heavily pruned, there is no guarantee that the new set can.  This could result in either FPGA resources being exhausted (admittedly less likely) or the latency drastically changing.  In such a failure, the only recourse is developing a new set of viable weights and hoping they succeed.  The time to do this is not predictable and certainly not welcomed if it delays operations. 
Using a dynamically loaded set of weights and biases will cost efficiency and the FPGA resources needed to support it may be greater but, since the FPGA image is unchanged and SNL is architected for deterministic behavior, this is a safe procedure with very predictable deployment times. It only has to be successively built once with neither the FPGA resources nor the latency changing.

When changes need to be made, many times it is in the face of multiple unrelated problems. Redeploying a new set of weights should not add to the problem list. Robustness in a real-time environment is part of good systems engineering.

\subsection{Why Streaming?} \label{stream}
A streaming interface connects the input of the current layer with the output of the previous layer. A memory interface delays the calculations of the current layer until all values of the previous layer are completed. The resulting latency accumulates though each layer using a memory interface.

In contrast, a \emph{streaming} interface allows the current interface to start its calculations as soon as the necessary data values are available, thus decreasing the latency.  In real-time applications, like triggering and feedback, latency is more valued than thru-put.  
The caveat is that some AI layer types are more amenable to streaming than others.  For example, a Conv2D layer can, depending on options, begin when roughly the number of rows and columns equal to 1/2 kernel dimensions are available.  Given most kernels are small, this delay will be small compared to the total data size. Other layers, such as the \emph{Dense} layers, can only output their first data value when \emph{all} the data have been processed.  Thus \emph{Dense} layers incur a heavy latency penalty.   While proper coding of such layers can provide high thru-put, no coding cleverness can avoid this latency penalty. This penalty should be taken into consideration when designing a low latency network. 

\subsection{Overview of SNL Usage}

The SNL user is presented as a collection of C++ templates that define the layer types and activators by specifying their parameters. In today's parlance, it is a header only package.

Current layers include among others, \emph{Conv2D, MaxPooling, AveragePooling, Dense, etc}.  The template parameters follow as closely as possible, in naming, ordering and meaning, the Python Keras methods for that layer.  Thus, users familiar with the Keras layer methods, should recognize their C++ template counterparts. This also has the upside that the very good documentation of the Keras layers can be referenced by the SNL user.

The user selects the layer and activator type and defines its parameters using the appropriate C++ templates. Where possible, sensible defaults are provided. These defaults are, by design, explicitly not hidden. This acts as a conspicuous prompt for the user to notice and change defaults when deemed necessary. 

Finally, in strictly a mechanical step and again with the philosophy of being as transparent to the network builder as possible, the layers are gathered, in the form of a simple list, by another C++ template to form the network.

\subsection{SNL Limitations, both Correctable and Intrinsic}
It is appropriate to be transparent about what SNL can and cannot do. 

First SNL is not a finished product. The basic architecture is sound but missing the following (ordered from the easier to harder to address)
   \begin{itemize}
       \item {Only a subset of all the Keras layers and activators are currently implemented}
       \item {Quantization of the weights and biases needs to be added}
       \item{Lack of global optimization across the network}
   \end{itemize}
   
 \subsubsection{New Layers and Activators:} Adding new layers and activators is tedious, but it is a well-defined procedure. This includes a defined testing and verification method when implementing new layers. Admittedly, the somewhat obscure syntax and style of C++ templates and meta-programming is off-putting. However, this is confined to the implementer who is expected to have the necessary skills and (considering the above critique of C++ syntax) the stamina to do this.  From the user's perspective, the resulting C++ templates are easy and straight-forward to use.  That is, the pain is confined to the few (the implementers), not the many (the users).
 
 \subsubsection{Quantization:} The quantization of the weights and biases has been shown to greatly reduce the latency and FPGA resource usage in AI inference engine implementations. Floating point are expensive in FPGAs. Quantization replaces these with the much cheaper arbitrary precision integers and scaled integers. As an extreme case, the literature includes implementations using 1-bit integers. Adding quantization is a matter of allocating the manpower and resources..
 
\subsubsection{Global Optimization:} The lack of global optimization is not as easily addressed as the above two. The streaming interface defined between layers is a form of global optimization, but there are other types that, at the level SNL is implemented, fundamentally cannot be. The balancing of FPGA latencies and resources across layers is only marginally addressed by judiciously specifying pragmas that, for example, unroll loops. Consider a layer that has minimal impact on the latency, but uses a disproportionate share of FPGA resources,\emph{e.g.} LUTs, DSPs.

A solution may be in a company Xilinx recently acquired and will soon be integrated into the HLS workflow.  The product, SLX, can be described as an post-processor to the FPGA synthesis stage.  Its promise is the user can specify global constraints on the resources including, not only logic resources, but also thru-put and latency. SLX will attempt to add pragmas that satisfies these constraints by considering the code in its entirety. How well this works in practice remains to be seen, but is an example of the needed solution.

\section{SNL For Convolution Networks}
This section is meant to give a flavor of the SNL implementation strategy using some of the layers typically found in a Convolution Network as examples. It also illustrates some of the challenges and techniques for implementing low-latency optimized code.  

\subsection{Data Widening}
A feature common, but not exclusive to Convolution Networks, is that many times, the input is a 3D tensor. In actual usage (\emph{i.e.} real hardware, delivering data in real-time) frequently two of these dimensions are presented serially, while the third dimension is parallelly available.  An example would be an RGB image.  The rows and columns are readout serially and the three colors in parallel.  The pattern is a number of sensors or channels each delivering distinct serial streams in parallel. A simple scheme would be to present each value as separate data items in the input serial stream.  Instead, SNL reflects the structure of the input, presenting all the channels in parallel, so instead of getting just a single data value in one FPGA clock, multiple values are fetched.  

In more than one network that has been implemented, the latency is a small number of fixed cycles associated with pipeline overheads plus a larger number of cycles proportional to the input data access time. Thus the time spent accessing the input data is often a significant contribution to the total latency, so handling this efficiently is important.  

The practice of widening the data path is very common in FPGA programming and fits naturally with Convolution Networks where often each channel of the initial layers is processed independently. Only latter, after the size of the data has been reduced by the initial layers, do these invoke a layer(s) that combines the channels. 

Of course there are practical limits on the width.  In FPGAs, a reasonable limit on the total width is 1-4K bits.  Thus, inputting 3 8 bit RGB values (24 bits total) or even 64 channels of 12 bit ADCs is permissible.  Given that the number of physical sensors/channels in a system is usually small, SNL currently assumes the third dimension can be always widen.  Clearly there will be exceptions and one of the challenges facing SNL is how to handle this.

\subsection{Controlling the Resource and Latency}
The selection of which FPGA used is often determined by 
 \begin{itemize}
     \item {Using a familiar FPGA family}
     \item {Given the cost of a FPGA, using the smallest one capable of meeting the requirements}
 \end{itemize}
 
 This translates into demanding the code squeezes as much performance using the fewest resources. High quality FPGA programming starts with selecting an algorithm and implementation that maps onto what an FPGA does well, then tuning the implementation, trying to find the sweet spot between performance and resource usage.
 
 SNL can help in the former, selecting and carefully coding the implementation of the layers and activators to be FPGA friendly, The latter, tuning the implementation, is a challenge to do in a user blind way. In HLS, specialized pragmas\footnote{Pragmas are a standard C/C++ feature used to communicate information directly to the compiler} are the vehicle that allows the mapping of the code to hardware resources.  It is through these pragmas that the performance/resource trade-off is realized.  Two common pragmas determine the amount of \emph{array partitioning} and \emph{loop unrolling}. Both affect performance and resource usage. A future strategy will be for SNL to provide reasonable defaults, but also a user accessible method to modify these if necessary.

\subsection{Scalability}
The above is one of a class of scalability problems. When programming a CPU, data array sizes and loop counts can be liberally increased with the only impact being execution time.  This is not true for FPGAs which have finite resources. So the challenge for SNL is how to handle cases when the finite size of an FPGA becomes a limitation. This came be summarized as
  \begin{itemize}
      \item {How much can and should SNL do \emph{under-the-hood}}
      \item {What control and how to expose that control to the user}
  \end{itemize}
Each path has its pitfalls.
   \begin{itemize} 
      \item{Can this truly be done without user input?}
      \item{Does giving the user tools to control this risk exposing details of the underlying implementation which, if the implementation needs to be modified, breaks user code?}
   \end{itemize}
   
\subsection{Activators} \label{activators}
The last step of an AI layer is the \emph{Activator}. SNL provides class templates for the common activators such as \emph{RELU}. From an implementation viewpoint these activators are divided into two orthogonal classes
   \begin{itemize}
       \item {Natural Floating Point}
       \item {One vs two pass}
   \end{itemize}
   
\subsubsection{Natural Floating Point:}
These are activators whose calculations are most naturally done in floating point and typically involve transcendentals, such as exponentials. An example would be the \emph{Sigmoid} activator.  These are computationally more expensive.  Work is needed to understand their usage when doing quantized integer implementations.  A look-up table would be a possible approach in this case. 

\subsubsection{One vs two pass:}
Many activators, \emph{e.g. Relu}, can process the data in a streaming fashion. Such activators are simple functions, when handed a data value, the function immediately returns a new value.  Some activators, \emph{e.g. SoftMax}, require two passes.  In the case of \emph{SoftMax}, the sum of all the values from the first-pass is used to normalize the output. 

Providing a standard interface for these is future SNL goal.  The interface should allow activators to be used interchangeably. The challenge will be for SNL to avoid a \emph{lowest common denominator} solution that unduly penalizes simple activators like \emph{Relu} just to accommodate the more involved two-pass types. 

However the reality is, independent of providing a clean interface, layers using two-pass activators are incapable of streaming.  Add to this that many of two-pass activators also involve floating point operations, makes them latency-cost expensive. The take-away is, similar to certain layer types, network designers should be aware of the unavoidable cost of using such activators.

\subsection{Layer Implementations: \emph{Conv2D}}

This section uses \emph{Conv2D} as a concrete example of a typical SNL layer implementation.  Two other layer types \emph{AveragePooling} and \emph{Dense} are used to illustrate other issues that occur.

\emph{Conv2D} is an almost complete implementation of the equivalent Keras method. The following gives a flavor of the correspondence between the C++ template and Keras method. Here is the Keras method's interface:

\begin{lstlisting}
keras.layers.Conv2D(filters, 
                    kernel_size,
                    strides              = (1, 1),
                    padding              = 'valid',
                    data_format          = None,
                    dilation_rate        = (1, 1),
                    groups               = 1,
                    activation           = None,
                    use_bias             = True,
                    kernel_initializer   = 'glorot_uniform',
                    bias_initializer     = 'zeros',
                    kernel_regularizer   = None,
                    bias_regularizer     = None,
                    activity_regularizer = None,
                    kernel_constraint    = None,
                    bias_constraint      = None)
\end{lstlisting}
This is the corresponding C++ template with the equivalent Keras parameter specification. 
\begin{lstlisting}
template<typename      SRC_STREAM,      -- data source

         size_t        NFILTERS,        -- filters
         
         size_t        KERNEL_NROWS,    -- kernel_size
         size_t        KERNEL_NCOLS,
         typename      KERNEL_TYPE,

         size_t        STRIDE_NROWS,    -- strides
         size_t        STRIDE_NCOLS,

         Padding       PADDING,         -- padding (Same or Valid)

         size_t        DILATION_NROWS,  -- dilation_rate
         size_t        DILATION_NCOLS,

         size_t        GROUPS,          -- groups

         typename      ACTIVATOR,       -- activator
         typename      BIAS_TYPE,

         typename      DST_TYPE,
         size_t        DST_AXIS_TID   = 0, 
         size_t        DST_AXIS_TDEST = 0>class Conv2D
\end{lstlisting}
The differences fall into two categories
  \begin{itemize}
      \item {General differences due to differences in Python and C++ syntax}
      \item {Differences in specifying specific parameters}
  \end{itemize}

\subsubsection{Parameter Defaulting:}Python, with its named parameters, as opposed to C++ templates' positionally based parameters, offers much cleaner defaulting. Having said that, by design SNL avoids, though not religiously, defaulting. 

Since defaults are specified in the interface, what the defaults are or even their existence is not immediately apparent when reading the code. This leaves the code vulnerable to changes in the defaults which may cause mysterious changes. SNL favors a bit more typing for the transparency it affords.

\subsubsection{Permissible Parameter Types:}
Python parameters can be any legitimate Python type.  C++ template parameters are limited to boolean, integer types and class types. In particular, floating point types are not permitted. To get around this, classes with purely constexpr's are used. Examples of this include the stream types (\emph{SRC\_STREAM, DST\_STREAM}) and the \emph{ACTIVATOR}. \footnote{Classes can be passed by reference.  This technique is used to make some of the weights and biases directly available at synthesis time in the \emph{Reservoir} layer.}

\subsubsection{Omitted parameters:}
The greatest noticeable difference is the absence of the \emph{xxx}\_initializer and \emph{xxx}\_constraint parameters. These are only used for the machine learning phase, so, having no use during the inference phase, are omitted in the C++ template.

\subsubsection{Data Types:}
Python deduces the data types of all the objects. For the most part this means a floating point type.  With quantization and the ability to specify a wider palette of data types (half-precision, arbitrary precision integers, \emph{etc.}), SNL must give control to the user via the \emph{XXX\_TYPE} parameters.

\subsubsection{SRC\_STREAM, DST\_STREAM:}
This specifies the source and destination data stream as a SNL template class. It can viewed as a combination of Python's numpy shape and a HLS stream \footnote{A HLS stream is the standard interface used to stream data between layers. It behaves like a FIFO. For technical reasons, the initial and final streams must be an AXI stream.}. 

Only the source stream for the initial layer needs to be defined by the user.  For subsequent layers, the source stream must be the previous layer's destination stream which can be easily referenced \footnote{The name of the class of the previous layer's destination stream is well-defined, for example, presuming the previous layer parameter definition is \emph{layer2::Parameters}, then \emph{layer2::Parameters::DstStream}.}. Furthermore, SNL can deduce the destination stream, with its stream shape being fixed by the layer type and its parameters. While the destination's stream data type can be overridden, it will default to that of the \emph{ACTIVATOR}.

\subsubsection{ACTIVATOR:} The activator is specified as class type. SNL provides templates for the common activators such as \emph{RELU}.  See the section \ref{activators} on activators for a more complete discussion.

\subsection{Layer Implementations: \emph{AveragePooling}}
The average pooling layer is very similar to \emph{Conv2D}. One could simply view this as a kernel with all entries equal to 1/Kernel\_Size. So, \emph{e.g.} if it were a 2x2 pool, then the entries would all be 1/4.  

The reason this is included in the discussion is that the division by the kernel size raises two important issues 
  \begin{itemize}
     \item {Division in FPGAs can be expensive in resources and time}
     \item {If using integer arthimetic, the division can result in bits being lost}
  \end{itemize}
  
 Because this is division by a fixed value, the usual trick of multiplying by precomputed reciprocal (if floating point) or by a binary scaled factor followed by removing the scale factor with a simply shift (if integer)  addresses the first issue.  However, if integer arithmetic, the loss of bits still remains.
 
 This is an issue SNL will have to be addressed when doing quantization.
 
\subsection{Layer Implementations: \emph{Dense}}
The dense layers are an example of a layer that kills streaming. By definition, all source data values must be available before an output value can be fully computed.  

The SNL implementation does calculate partial results based on the available source data. This helps increase the thru-put and at least minimize the latency, but the minimum baseline latency is set by the number of source data values. As stated previously, no amount of clever coding can be avoid this penalty. One practical consideration helps; \emph{Dense} layers generally occur as the final layers in a network when the data sizes have been reduced. The only other option available is to avoid using layers similar to the \emph{Dense} layers.

\subsection{Adding New Layers and Activators} \label{add_new}
In implementing SNL, a standard prescription for defining new layers and activators has slowly developed. While, making no claims that it is easy, this prescription is well defined.  A suite of support functions and template classes are available to assist.  At the risk of becoming too meta, SNL now becomes not only a set templates to define an arbitrary network, but also a set of rules with compile-time support methods for extending the palette of layers and activators. To use an overused phrase, this make SNL more amenable to open sourcing.

\subsubsection{Example of a support method:}
As an illustration of how similar writing a new layer is to writing a compiler, consider an implementation for a 3x3 2D convolution using arbitrary precision integers. This involves multiplying and adding 9 data values by 9 kernel weights, \emph{i.e.} a common dot product.  The question is: What is the data type of the dot product sum?

The compiler can easily determine the size of each multiplication. It is simply the sum of the number of bits of the two multipliers.  However the data type of the sum must be wide enough to avoid overflows.  In usual CPU destined code, this is solved by \emph{overkill}; either doing things in floating point or using very wide integers. However, to minimize resource costs in an FPGA, the type and size of the data types should be kept at a minimum.

The compiler is not equipped to do this. To help, SNL uses meta-programming features of C++ to the define a compile time method that takes the two input types and the number of summed elements and returns the appropriate data type.

\begin{lstlisting}
   snl::datatype::DotType<Type0, Type1, Count>
\end{lstlisting}

For example, consider a 3x3 kernel of 8-bit signed integer weights with 12-bit unsigned data types.  The minimum output data type of this convolution is a 24 bit signed integer, 20 bits from the multiple and 4 bits to cover the 9 sums. Using the above

\begin{lstlisting}
   // Define the dot product type
   using DotProduct_t = snl::datatype::DotType(ap_uint<12>, 
                                               ap_int<8>,
                                               9>;
   // Do a (overly) simple implementation of a dot product
   DotProduct_t sum = 0.
   for (int i = 0; i < 9; i++) sum += kernel[i] * data[i];
\end{lstlisting}

\section{Implementation Results: BES Network}
\begin{figure}[htp]
    \centering
    \includegraphics[width=\textwidth]{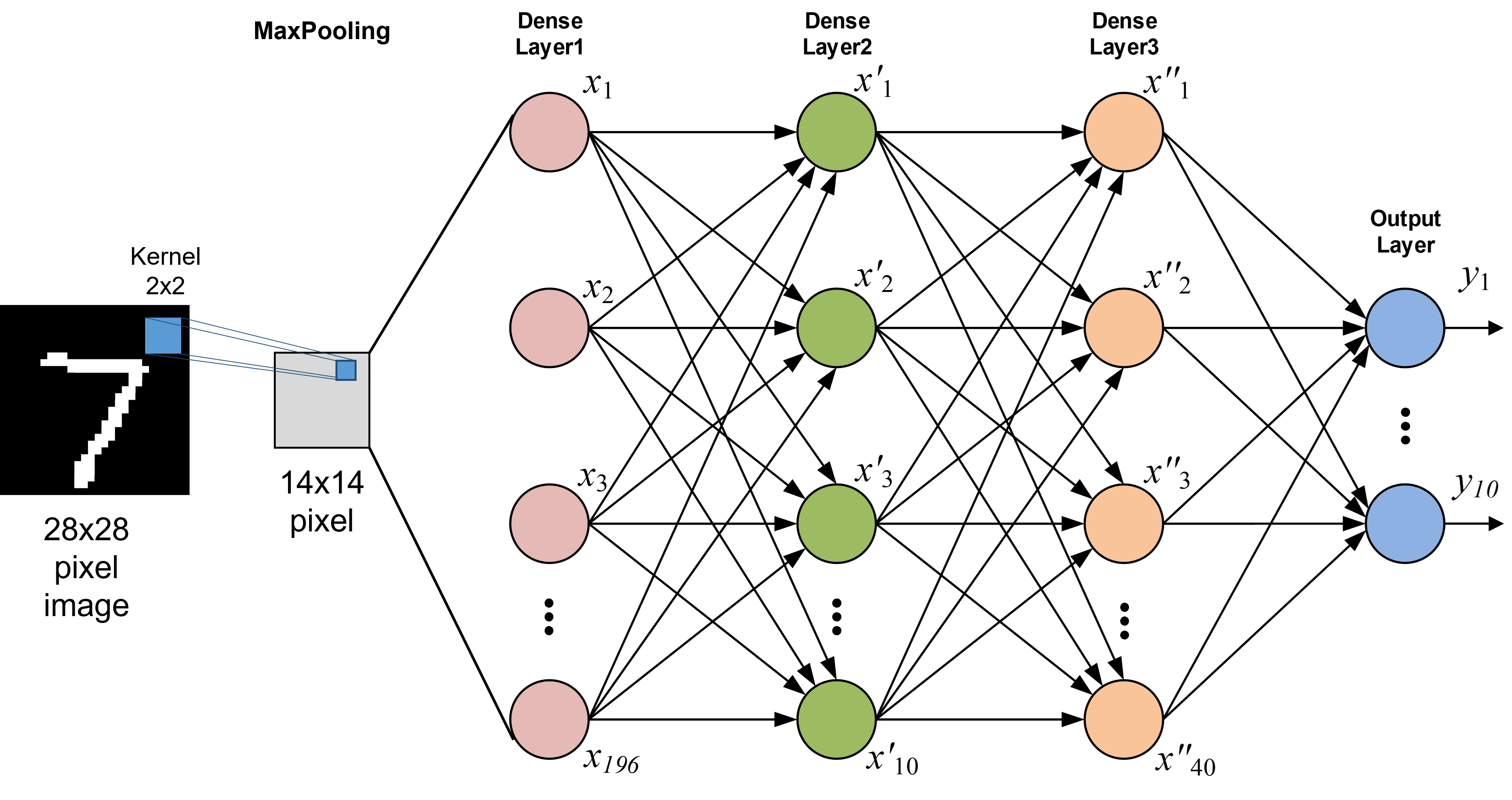}
    \caption{BES network model}
    \label{fig:1}
\end{figure}
\begin{table}[ht!]
\centering
\begin{tabular}{|c|c|c|} 
 \hline
 Layer(type) & Output Shape & Activator \\ [0.5ex] 
 \hline
 Maxpooling2D & 14x14 & -          \\ 
 Flatten      & 196   & -          \\
 Dense        & 10    & Leaky ReLU \\
 Dense        & 40    & Leaky ReLU \\
 Dense        & 10    & Leaky ReLU \\
 \hline
\end{tabular}
\caption{BES network layers.}
\label{table:1}
\end{table}

The BES network consists of one MaxPooling2D layer and three dense layers with the Leaky-ReLU activation function. In this paper, the MNIST dataset is used to test and verify the BES network. The configuration of the network is described in Fig. 1 and Table 1. The weights and biases of each layer are calculated using the Keras and extracted for the SNL framework. The framework is verified by comparing the result of the network output with the Keras.\hfill 

We were able to compile the BES network into a Xilinx KCU1500 device using the Xilinx HLS synthesis followed by a separate place and route with the HLS output included as a module in a larger VHDL based design. We used a python based software framework to load the FPGA, configure the weights and bias and to DMA image data into the FPGA in a streaming fashion. The inference results are then received via DMA using this same software. 

Our python framework is able to read in the weight and bias data generated by the Keras tool directly and then formats this data to match the memory layout in the FPGA. Similarly the image data itself is read using this tool in its native format and streamed into the Xilinx FPGA. The FPGA results are stored in a custom data file and later compared to the results set generated by Keras. Our testing found perfect correlation between the results generated by Keras and the results received from the firmware in the Xilinx KCU1500. 

The table 2 outlines the resource usage of the compiled BES network module:

\begin{table}[ht!]
\centering
\centering
\begin{tabular}{|c|c|c|} 
 \hline
Resource & Usage & KCU1500 Total \\ [0.5ex] 
 \hline
DSPs & 298 & 5,520 \\
FFs & 36,901 & ~1M  \\
LUTs & 26,016 & 663,360  \\
BRAM & 38 (0.7MB) & 75Mb  \\
\hline
\end{tabular}
\caption{BES network Ulilization.}
\label{table:2}
\end{table}

In its current use the network runs at a clock rate of 250Mhz. Based upon compilation results be believe this can be raised higher possibly to 300Mhz or more. Further testing will allow us to determine how fast we can run this network. Given a clock rate of 250Mhz we were able to achieve a total latency for each image frame's inference of 1.1015uS. This is well within the target requirement for the initial application we are targeting.

The results reflect the fact that when moving an image from a dense layer to a dense layer in our current implementation does not include a pipelining stage. This saves memory in that it does not introduce additional registers at the output of each computation engine. We plan to allow a flag which will enabling the user to determine if they want to introduce pipelining at each dense layer output. The advantage of a pipeline is that it would allow a frame interval which is less than the total inference latency of the network. This will be required for larger inference networks running at higher frame rates.

\section{SNL For Reservoir Networks}
The Reservoir Network is included here because it uses a layer that is fairly different from other layers in the following ways
  \begin{itemize}
      \item {It carries state information in the form of the previous reservoir vector}
      \item {It has two classes of weights}
         \begin{itemize}
            \item{a fixed, built-in set.  They are randomly initialized in the training phase, but they themselves are not trained.}
            \item{a traditional set of weights and biases that are trained and downloaded}
          \end{itemize}
  \end{itemize}
  
  The other salient feature is that, because it connects all neurons, it cannot be streamed.  An example network using 2 \emph{Reservoir} layers has been implemented. A feature of these networks is the reservoir vector is large, in the 1000s.  The weight matrix connecting all the neurons of the reservoir network is thus this dimension squared.  This results in a very large memory to hold it and a corresponding large number of DSPs to achieve the necessary low-latency.  Such networks will need a correspondingly large FPGA.  In the Xiinx series of FPGAs, choosing one with URAM is advantageous.  
  
  Again because of the large sizes involved, whereas SNL's typical target network latency is \textasciitilde 10 \emph{usecs}, realistic Reservoir networks are in \textasciitilde .1-10 \emph{msecs}.

\section{Conclusion}
While SNL is still a work in progress, it has shown promise in implementing very low-latency networks.  These latencies are very close to the best that can be expected based on the input data sizes and the unavoidable pipeline delays of the layers. The data does nicely stream through the layers.

In the future we intend to continue to add the necessary new layers and activators as stated in this document and study how their used impacts resource utilization and latency of the networks they are used in. We also plan to start implementing quantization in the models, including some novel techniques being studied by other groups such as non linear quantization. Our key concern here is to ensure we handle the overflow cases in a way which preserves the accuracy of the inference and does not introduce additional complexity in the activators. 

Also now that we have a baseline framework in place we want to compare the resource usage of this framework with other frameworks that are available. Previous experience with these other frameworks indicates that they do not scale well to the network sizes we are looking at, but we intend to do direct comparisons of various size networks, starting small and scaling up to larger networks to see where the various frameworks no longer scale and how the implementation latencies compare.

\printbibliography

\end{document}